\documentclass[submission,copyright,creativecommons]{eptcs}
\providecommand{\event}{ICLP 2021} 

\usepackage{mathptmx}

\usepackage{url}  
\usepackage{graphicx}  
\usepackage{xspace}
\usepackage{xcolor}
\usepackage{enumitem}
\usepackage[mathscr]{euscript}
\usepackage{comment}
\usepackage{amsmath}
\usepackage{framed}

\def\verb#1{\hbox{\tt #1}\xspace}

\def\apia{$\mathscr{APIA}$\xspace}
\def\aia{$\mathscr{AIA}$\xspace}
\def\aopl{$\mathscr{AOPL}$\xspace}

\def\al{$\mathscr{AL}$\xspace}

\newtheorem{definition}{Definition}

\title{$\mathscr{APIA}$: An Architecture for Policy-Aware Intentional Agents}
\author{John Meyer
\institute{Miami University, Ohio, USA}
\email{meyerjm@miamioh.edu}
\and
Daniela Inclezan
\institute{Miami University, Ohio, USA}
\email{inclezd@miamioh.edu}
}

\def\titlerunning{$\mathscr{APIA}$: An Architecture for Policy-Aware Intentional Agents}
\def\authorrunning{J. Meyer \& D. Inclezan}
\begin{document}
\maketitle

\begin{abstract}
This paper introduces the \apia architecture for policy-aware intentional agents. These agents, acting in changing environments, are driven by intentions and yet abide by domain-relevant policies. This work leverages the \aia architecture for intention-driven intelligent agents by Blount, Gelfond, and Balduccini. It expands \aia with notions of policy compliance for authorization and obligation policies specified in the language \aopl by Gelfond and Lobo. \apia introduces various agent behavior modes, corresponding to different levels of adherence to policies. \apia reasoning tasks are reduced to computing answer sets using the \textsc{clingo} solver and its Python API.
\end{abstract}

\section{Introduction}
\label{sec:intro}
This paper introduces \apia,\footnote{\apia stands for ``Architecture for Policy-aware Intentional Agents.''} 
an architecture for intentional agents that are aware of policies and abide by them.
It leverages and bridges together research by Blount, Gelfond, and Balduccini \cite{bj13,bgb14}
on intention-driven agents (\aia\footnote{\aia stands for ``Architecture for Intentional Agents.''}), and work by Gelfond and Lobo \cite{gl08} 
on authorization and obligation policy languages (\aopl\footnote{\aopl stands for ``Authorization and Obligation Policy Language.''}). Both \aia and \aopl are expressed in
action languages \cite{gl98} that have a 
seamless translation into 
logic programs \cite{gl91},
and are implemented in Answer Set Programming (ASP) \cite{mt99}.

With the current rise in autonomous systems 
the question arises of how to best construct agents that are capable of acting in a changing environment 
in pursuit of their goals, while also abiding by domain-relevant policies. For instance,
we would want a self-driving car to not only take us to our destination but also do so while respecting
the law and cultural conventions. 
This work is a first step in this direction. 
As in Blount {\em et al.}'s work, we focus on agents: 
\begin{itemize}
\item whose environment (including actions and their effects) and mental states 
can be represented by a transition diagram. Physical properties of the environment and 
the agent's mental states form the nodes of this transition diagram. Arcs from one state to another are
labeled by actions that may cause these transitions.
\item who are capable of making correct observations, remembering
the domain history, and correctly recording
the results of their attempts to perform actions;
\item who are {\em normally} capable of observing the occurrences
of exogenous actions; and 
\item whose knowledge bases may contain some pre-computed plans for achieving certain goals,
which we call {\em activities}, while agents can compute other plans on demand.
\end{itemize}
Additionally, these agents possess specifications
of authorization and obligation policies and have access to reasoning mechanisms for
evaluating the policy compliance of their actions.

Thus, in our work, we extend the \aia architecture introduced by Blount {\em et al.} \cite{bj13,bgb14}
with the notion of policy compliance. \aia agents are Belief-Desire-Intention (BDI) agents \cite{rg91} who are 
driven by goals, have intentions that persist,
and can operate with pre-computed activities. However, they are oblivious to the different nuances of
authorization and obligation policies (e.g., actions whose execution is permitted vs. not permitted), 
and policy compliance. These intentional agents have only two
behavioral modes: either ignore policies altogether 
(when policies are not represented by any means in the agent's knowledge base)
or blindly obey policies that disallow certain actions from being performed, 
even if this may be detrimental
(when policies are represented as actions that are impossible to be executed). 
Instead, we introduce a wider range of possible behaviors that may be set by an agent's controller
(e.g., prefer plans that are {\em certainly} policy-compliant to others 
that are only {\em possibly} policy-compliant, but if no
plans of the former type exist, accept plans of the latter kind).

In formalizing the different policy compliance behavior modes, 
we rely on the language \aopl \cite{gl08} for specifying authorization and obligation policies,
and priorities between such policies. While Gelfond and Lobo specify reasoning algorithms for determining the
degree of policy compliance for a sequence of actions, this is done from the perspective of a third-person observer
analyzing the actions of {\em all} agents {\em after} they already occurred. 
Instead, our work focuses on an agent making decisions
about its own future actions. As a result, various courses of action may be compared to determine the most
policy-compliant one according to the behavior mode set by the agent's controller. 
Moreover, our research addresses interactions between authorization and obligation policies that were not discussed
in the work on \aopl.

\smallskip
There have been several other attempts to enable agents to reason over various kinds of policies.
However, these involve reasoning over access control policies only \cite{fck95,so16,af17} 
and a few utilize ASP as a reasoning tool for this purpose \cite{bs12,bbgg14}.
Access control policies are more restrictive than the kinds of policies an agent using \aopl can reason over.
The work that is closest to our goal is the PDC-agent by Liao, Huang, and Gao \cite{lhg06}.
The PDC architecture extends the BDI architecture with a policy and contract-aware methodology the authors call 
BGI-PDC logic. A PDC-agent is an event-driven multi-component framework which allows for controlled and 
coordinated behavior among independent cooperative agents.
Liao {\em et al.} use policies to control agent behavior, and contracts as a mechanism to coordinate actions between agents.
This architecture was later extended to support reasoning over social norms (the NPDC-agent) \cite{pgawg08}.
The PDC-agent architecture is defined as a 7-tuple of the following components: (Event Treating Engine, Belief Update, Contract Engine, Policy Engine, Goal Maintenance, Plan Engine, Plan Library)~\cite{lhg06}.
A major distinction of the PDC-agent agent architecture is that it supports coordination among multiple agents. 
This is beyond the scope of our work. Both \aia and \apia focus on an agent with individual goals. 
Expanding these architectures into multi-agent frameworks by introducing communication acts is still part of future work.
However, knowledge about the changing environment is expressed in the PDC-agent architecture 
in terms of a Domain Conceptualization Language (DCL)~\cite{gyw05} and a Concept Instance Pattern (CIP).
While DCL and CIP can represent plans (which are analogous to activities in the \aia architecture), 
there is no support for expressing the direct or indirect effects of an action.
This is a disadvantage in comparison to action-language-based architectures since plans have to be 
pre-computed and the goals that they accomplish must be annotated according to the agent's designer's intuition.
Since action languages only require a description of the effects of individual actions 
(and plans consisting of all combinations of actions can be automatically computed), 
there is significantly less work for a human designer when working with \apia than the PDC-agent architecture.

{\em Thus, our proposed \apia architecture is, to the best of our knowledge, 
the only intentional agent architecture that is capable to model compliance with
complex authorization and obligation policies, while allowing agents to come up with
policy-compliant activities on the fly.}

\smallskip
The major contributions presented in this paper work are as follows:
\begin{enumerate}
\item Create a bridge between research on intentional agents and policy compliance, 
thus producing a {\em policy-aware intentional agent} architecture \apia.
\item Introduce various agent behavior modes with respect to compliance with authorization and obligation policies.
\item Introduce mechanisms to check the consistency of a policy containing both authorization and obligation policies
and reason over the interactions between these two.
\item Implement \apia in \textsc{clingo} (version 5.4.1)\footnote{\url{https://potassco.org/clingo/}} while leveraging \textsc{clingo}'s Python API.\footnote{\url{https://potassco.org/clingo/python-api/5.4/}}
(As a by-product, \aia was also updated to this \textsc{clingo} version).
\end{enumerate}

\section{Background}
\label{sec:background}

In this section, we briefly present the \aia architecture and \aopl language, which 
form the two pillars of our work. 
We direct the unfamiliar reader to outside resources on Answer Set Programming \cite{gl91,mt99} and
action language \al \cite{gk14}, which are also relevant to our research.

\subsection{\texorpdfstring{\aia}{AIA}: Architecture for Intentional Agents}

\label{subsec:aia_architecture}

The \textit{Architecture for Intentional Agents}, \aia \cite{bgb14}, builds upon the Observe-Think-Act control loop
of the AAA architecture\footnote{AAA stands for ``Autonomous Agent Architecture.''} \cite{bg08} and extends it in a couple of directions. 
First, \aia adds the possibility for action failure:
the agent \textit{attempts} to perform an action during its control loop but may find that it is unable to do so.
In this case, the action is deemed \textit{non-executable}.
Second, \aia addresses a limitation of AAA in which plans are not persisted across iterations of the control loop.
In \aia, agents pursue goals by persisting in their intentions to execute 
action plans known to satisfy these goals (i.e., activities).
Activities are represented via a set of statics~\cite{bj13,bgb14}:
$$
\begin{array}{l}
		\{ activity(M), \ \ length(M, L), \ \ goal(M, G),\\
    \ \ component(M, 1, C_1), \ \ component(M, 2, C_2), \ \ \dots, \ \ component(M, L, C_L)\ \} \\
\end{array}
$$
where $M$ is a unique identifier for the activity; 
$C_1, C_2, \dots, C_L$ 
are the $1^{st}, 2^{nd}, \dots, L^{th}$ \textit{components} of the activity; 
and $G$ is the goal that $[C_1,C_2,\dots,C_L]$ achieves.
Some activities are pre-computed and stored in the agent's knowledge base, while the rest can be generated on demand.

In addition to fluents and actions describing the agent's environment, 
\aia introduces \textit{mental fluents} to keep track of the agent's progress in the currently intended activity
and towards its desired goal.
Mental fluents are updated through \textit{mental actions}.
The Theory of Intentions 
is a collection of axioms that maintain an agent's mental state.
Elements of the agent's mental state include the currently selected goal $G$, 
stored in the $active\_goal\left(G\right)$ inertial fluent, and the current planned activity, stored in the $status(M, k)$ inertial fluent.
When either a goal is selected or an activity is planned, they are said to be \textit{intended}.
Mental action $start(M)$ initiates the agent's intention to execute activity $M$ and $stop(M)$ terminates it.
Though most actions are executed by the agent itself, some must be executed by the agent's controller.
The exogenous mental action $select(G)$ causes the agent to intend to achieve goal $G$ while
$abandon(G)$ causes the agent to cease its intent to achieve goal $G$.

An agent in the \aia architecture performs the following loop:
\begin{equation}
\label{eq_loop}
\begin{array}{l}
\mbox{1. Interpret observations.}\\
\mbox{2. Find intended action } A.\\
\mbox{3. Attempt to perform } A\mbox{; record this attempt in history.}\\
\mbox{4. Observe the world; record observations in history.}\\
\mbox{5. Repeat (i.e.~go to step 1).}
\end{array}
\end{equation}

\subsection{\texorpdfstring{\aopl}{AOPL}: Authorization and Obligation Policies in Dynamic Systems}

In real-world applications, an autonomous agent may be required to follow certain rules or ethical constraints, and 
may be penalized when acting in violation of them.
Thus, it is necessary to discuss policies for agent behavior and a formalism with which agents can deduce the compliance of their actions.
Gelfond and Lobo \cite{gl08} introduce the Authorization and Obligation Policy Language \aopl for policy specification.
An {\em authorization} policy is a set of conditions that denote whether an agent's action is permitted or not.
An {\em obligation} policy describes what an agent must do or must not do.
\aopl works in conjunction with a dynamic system description written in an action language such as \al.
An agent's policy is the subset of the trajectories in the domain's transition diagram 
that are desired by the agent's controller.

Policies of \aopl are specified using predicates $permitted$ for authorization policies,
$obl$ for obligation policies, and static laws similar to those from action language \al:
$$
\begin{array}{rcl}
        permitted\left(a\right) & \textbf{ if } &cond \\
        \neg permitted\left(a\right) &\textbf{ if } &cond \\
				obl\left(h\right) &\textbf{ if } &cond \\
        \neg obl\left(h\right) &\textbf{ if } &cond
\end{array}
$$
where $a$ is an action; $h$ is a happening (i.e., an action or its negation\footnote{If $obl(\neg a)$ is true, 
then the agent must not execute $a$ in the current state.}); and
$cond$ is a, possibly empty, conjunction of fluents, actions, or their negations.
In addition to these strict policy statements, \aopl supports defeasible statements and 
priorities between them as in:
\begin{equation}
\label{eq2}
\begin{array}{lccl}		
        d: \textbf{normally } & permitted(a) & \textbf{ if } & cond \\
        d: \textbf{normally } & \neg permitted(a) &\textbf{ if } &cond\\
        d: \textbf{normally } & obl(h) &\textbf{ if } &cond \\
        d: \textbf{normally } & \neg obl(h) &\textbf{ if } &cond \\
				& prefer(d_i, d_j)&&
\end{array}
\end{equation}

Gelfond and Lobo define {\em policy compliance} separately for authorizations vs. obligations:
\begin{definition}[Policy Compliance -- adapted from Gelfond and Lobo \cite{gl08}]
\label{def1}
{\bf Authorization}: A set $A$ of actions occurring at a transition system state $\sigma$ 
is \textit{strongly compliant} with a policy $P$ if all $a \in A$ are known to be permitted at $\sigma$ using rules in $P$.
$A$ is \textit{non-compliant} with $P$ if any of the actions are known to be not permitted at $\sigma$ using rules in $P$.
Otherwise, if $P$ is unclear and does not specify whether an action $a \in A$ is permitted or not at $\sigma$, 
then the set of actions is \textit{weakly compliant}.

\noindent
{\bf Obligation}: A set of actions $A$ occurring at state $\sigma$ 
is {\em compliant} with an obligation policy $P$ if $a \in A$ whenever $obl(a)$ 
can be derived from the rules of $P$ and $a \not \in A$ whenever $obl(\neg a)$ can be derived from $P$ at $\sigma$.
Otherwise, it is {\em non-compliant}.
\end{definition}
Note that a set of actions $A$ can be strongly, weakly, or non-compliant with an authorization policy, 
but $A$ can only be compliant or non-compliant with an obligation policy.
Given an \aopl policy (with authorization and obligation policy statements), $A$ is \textit{strongly compliant} 
if it is strongly compliant with its authorization policy and compliant with its obligation policy.
Likewise for \textit{weak compliance} and \textit{non-compliance}.
Computing policy compliance is reduced to the problem of finding answer sets of a logic program
obtained by translating the policy rules into ASP. In this translation, predicates $obl$ and $permitted$
are extended to include an extra argument $I$ standing for the time step, while $lp(x)$ denotes
the ASP transformation of $x$, where $x$ can be a rule, an action literal, a fluent literal, or a condition.

\aopl does not discuss interactions between authorization and obligation policies on the same action,
does not define compliance in terms of obligation policies for a trajectory in the dynamic system,
and does not compare the degree of compliance of two trajectories. All of these aspects are relevant when modeling
a policy-aware intentional agent and are addressed in our work.


\section{\texorpdfstring{\apia}{APIA} Architecture}

We can now introduce our \apia architecture for policy-aware intentional agents.
We focus on two main aspects of \apia: reframing \aopl to fit an agent-centered architecture,
and the encoding of different policy compliance behavior modes of an \apia agent.

\subsection{Re-envisioning \texorpdfstring{\aopl}{AOPL} Policies in an Agent-Centered Architecture}

Gelfond and Lobo \cite{gl08} conceived \aopl as a means to evaluate policy compliance in a dynamic system.
This differs from \aia in the following ways:

\begin{itemize}
    \item \aopl evaluates trajectories in a domain's transition diagram from a global perspective whereas \aia distinguishes between agent actions and exogenous actions, and chooses which agent actions to attempt next.
    \item \aopl evaluates histories at ``the end of the day'' whereas the \aia architecture, while still reasoning over past actions in its diagnosis mode, places an emphasis on planning future actions to achieve a future desired state.
\end{itemize}

These differences prevent \aopl policies from interoperating with the \aia agent architecture out of the box.
To address the first issue, we constrain \aopl policies to describe only agent actions in our \apia architecture.
For the second issue, we adjust our policy compliance rules such that only future actions affect policy compliance.
Since our focus is on planning, in \apia past actions are always considered ``compliant'' although they might not have been at the time.
For an agent that previously had no choice but a non-compliant action, this allows the agent to conceive of ``turning a new leaf'' and seeking policy-compliant actions in the future.

Also, \aopl does not include specification on how authorization policy statements interact with obligation policy statements.
For example, consider the following \aopl policy:
$$
\begin{array}{c}
    permitted(a) \\
    obl(\neg a)
\end{array}
$$
which is contradictory, since the agent is permitted to perform action $a$ but at the same time is obligated to refrain from it.
Appealing to common sense, if an agent is obligated to refrain from an action, one would conclude that the action is not permitted.
Likewise, it makes sense to say that, if an agent is obligated to do an action, then it must be permitted.
Thus, we take these intuitions and create the following non-contradiction ASP axioms, in which we use literals
$obl$ and $permitted$ expanded to include a new argument $I$ representing the time step:
\begin{equation}
\label{eq0}
\begin{array}{ll}
\leftarrow & agent\_action(A),\ \  obl(A, I),\ \  \neg permitted(A, I).\\
\leftarrow & agent\_action(A), \ \ obl(neg(A), I), \ \ permitted(A, I).
\end{array}
\end{equation}
These enforce that, at the very least, the authorization and obligation policies do not contradict each other,
while allowing for defeasible policies to work appropriately.

We also extend the translation of defeasible policy statements.
Suppose we have the following:
\begin{equation}
\label{eq1}
\begin{array}{c}
    \textbf{normally } permitted(a) \\
    obl(\neg a) \ \textbf{ if } \  cond
\end{array}
\end{equation}
Using Gelfond and Lobo's approach \cite{gl08}, the corresponding ASP translation would be:
$$
\begin{array}{l}
    permitted(a, I) \leftarrow \textbf{not } \neg permitted(a, I). \\
    obl(neg(a), I) \leftarrow lp(cond).
\end{array}
$$
where $lp(cond)$ and $obl(neg(a), I)$ represent the logic programming encoding of $cond$ and $obl(\neg a)$, respectively.
Based on policy (\ref{eq1}), both $permitted(a, I)$ and $obl(neg(a), I)$ would be true at a time step $I$ when $cond$ is met.
This violates the non-contradiction axioms in (\ref{eq0}).
So, we replace the translation of the defeasible statement in (\ref{eq1}) 
with the following encoding:
$$
\begin{array}{lll}
    permitted(a, I) & \leftarrow & \textbf{not } \neg permitted(a, I), \ \ \textbf{not } obl(neg(a), I).
\end{array}
$$
This allows the presence of $obl(\neg a)$ to be an exceptional case to the defeasible rule.

In general, we propose translating the different types of defeasible statements in (\ref{eq2})
as follows, respectively, where predicate $ab$ facilitates dealing with possible additional (weak) exceptions:
$$
\begin{array}{rll}
permitted(a, I) & \leftarrow & lp(cond), \ \ \textbf{not } ab(d, I), 
        \ \ \textbf{not } \neg permitted(a, I), \ \ \textbf{not } obl(neg(a), I).\\
\neg permitted(a, I) & \leftarrow & lp(cond), \ \ \textbf{not } ab(d, I), 
				\ \ \textbf{not } permitted(a, I), \ \ \textbf{not } obl(a, I).\\
obl(a, I) & \leftarrow & lp(cond), \ \ \textbf{not } ab(d, I), 
				\ \ \textbf{not } \neg obl(a, I), \ \ \textbf{not } \neg permitted(a, I).\\
obl(neg(a), I) & \leftarrow & lp(cond), \ \ \textbf{not } ab(d, I), 
				\ \ \textbf{not } obl(neg(a), I), \ \ \textbf{not } permitted(a, I).								
\end{array}
$$

\subsection{Policy-Aware Agent Behavior}


The \aia architecture, which is the underlying basis of \apia, introduces {\em mental} fluents and actions
in addition to physical ones.
In \apia, we additionally introduce {\em policy} fluents and actions needed to reason over policy compliance
(see Table \ref{table:policy_fluents_and_actions}).
The new policy action descriptions encode the effects of future agent actions on policy compliance and 
provide means for the control loop to deem non-compliant activities futile and 
execute compliant ones in their place.

\begin{table}[!hbt]
\caption{List of Policy Fluents and Actions in the \apia Architecture}
\label{table:policy_fluents_and_actions}
\begin{minipage}{\textwidth}
\begin{tabular}{l l |l}
\hline\hline
{\bf Fluents} & & {\bf Actions} \\
  \hline
	{\bf Inertial:} & $auth\_compliance(strong)$  & $ignore\_not\_permitted(a)$\\
	                & $auth\_compliance(weak)$    & $ignore\_neg\_permitted(a)$\\
									& $obl\_compliant(do\_action)$ & $ignore\_obl(a)$\\
									& $obl\_compliant(refrain\_from\_action)$ & $ignore\_obl(neg(a))$ \\
	{\bf Defined:}  & $policy\_compliant(f)$, for every physical fluent $f$ \ \ \ & \ \ \ \ for every physical action $a$\\
\hline\hline
\end{tabular}
\end{minipage}
\end{table}

For example, the dynamic causal laws in (\ref{eq5}) define inertial policy fluents
$auth\_compliance(weak)$ and $auth\_compliance(strong)$ according to the 
definitions for strong and weak authorization policy compliance of \aopl seen in Definition \ref{def1}:
\begin{equation}
\label{eq5}
\begin{array}{lllll}
    a & \textbf{ causes } &\neg auth\_compliance(strong) & \textbf{ if } & \textbf{ not } permitted(a) \\
    a & \textbf{ causes } & \neg auth\_compliance(weak) & \textbf{ if } & \neg permitted(a)
\end{array}
\end{equation}
These rules are defined for every physical action $a$ of the transition system.
Should an action $a$ occur where $permitted(a)$ is not known to be true, 
then the scenario ceases to be strongly compliant (i.e., it becomes weakly compliant).
Since $auth\_compliance(strong)$ cannot be made true again by any action, the rest of the scenario remains weakly compliant by inertia.
Likewise, should an action $a$ occur where $permitted(a)$ is false, then the scenario ceases to be weakly compliant 
(i.e., it becomes non-compliant) and remains in this state by inertia.

For every physical fluent $f$, we introduce a new defined policy fluent $policy\_compliant(f)$.
This allows us to reuse the \aia control loop shown in (\ref{eq_loop}) as is.
When the agent controller wants to specify that the agent should achieve goal $f$
in a policy-compliant manner, the controller simply has to initiate action $$select\_goal(policy\_compliant(f))$$
instead of the original $select\_goal(f)$.
The policy fluent $policy\_compliant(f)$ is true iff $f$ is true and $auth\_compliance(l)$ is true, 
for some minimum compliance threshold $l$ set by the controller.
Thus, when $policy\_compliant(f)$ is an agent's goal, activities below $l$-compliance are deemed as futile 
and the agent works to achieve fluent $f$ subject $l$-compliance.

\medskip
\noindent
{\bf Authorization Policies and Agent Behavior. }
To allow for cases when the threshold $l$ set by the controller is not the maximum possible level of compliance
or, in the future, cases when an agent deliberately chooses to act without $l$-compliance, 
we add policy actions 
$ignore\_not\_permitted(a)$ and $ignore\_neg\_permitted(a)$, where $a$ is a physical action.
By executing these actions concurrently with $a$, 
our agent ignores $a$'s effect at that time step on weak compliance or non-compliance, respectively.
This enables our agent to look for activities with a lower level of compliance if no activities that achieve
$f$ are strongly-compliant.
One can imagine that this capability can be used to model multiple agent behaviors, 
based on what the minimum and 
maximum requirements of adherence to their authorization policy are.
We have parameterized the agent's behavior as seen in Table \ref{table:apia_authorization_modes},
and introduced names for these possible agent behaviors.

\begin{table}[!hbt]
    \centering
    \begin{tabular}{  l | c | c | c  }
        \hline\hline
        & {\bf Require weak} & {\bf Prefer weak over non-compl.} & {\bf Ok with non-compl.} \\
        \hline\hline
        {\bf Require strong} & Paranoid & (Invalid) & (Invalid) \\
        \hline
        {\bf Prefer strong over weak} & Cautious & Best effort & (Invalid) \\
        \hline
        {\bf Ok with weak} & Subordinate & Subordinate when possible & Utilitarian \\
        \hline\hline
    \end{tabular}
    \caption{\apia Authorization Policy Modes}
    \label{table:apia_authorization_modes}
\end{table}

One behavior mode is for the agent to strictly adhere to its authorization policy such that it never 
chooses to perform $ignore\_neg\_permitted(a)$.
This causes all non-compliant actions to indirectly cause $policy\_compliant(f)$ to be false, if they are executed.
Hence, only activities with weakly or strongly compliant actions are considered.
Since this mode never dares to become non-compliant, it is called \textit{subordinate}.

A similar behavior mode causes the agent to never perform $ignore\_not\_permitted(a)$.
This causes all weak and non-compliant actions to indirectly cause $policy\_compliant(f)$ to be false when executed.
Hence, only activities with strongly compliant actions are considered.
Since weakly compliant actions are actions for which the policy compliance is unknown, 
this mode is called \textit{paranoid} as it treats weakly compliant actions as if they were non-compliant.

Another behavior mode allows unrestricted access to the two actions, 
$ignore\_neg\_permitted(a)$ and $ignore\_not\_permitted(a)$.
This mode is called \textit{utilitarian} because it reduces the behavior of \apia to that of \aia, 
where policies are not considered at all.

\smallskip
An interesting feature of the $ignore\_neg\_permitted(a)$ and $ignore\_not\_permitted(a)$ actions is the ability 
to optimize compliance.
Using preference statements in ASP, we can require the control loop to minimize the use of 
these two policy actions.
Hence, if it is possible to execute an activity that is strongly compliant, the agent will prefer it over a weakly or 
non-compliant one (since the use of these actions is required to allow $policy\_compliant(f)$ to be true).
Under this condition, $ignore\_not\_permitted(a)$ and $ignore\_neg\_permitted(a)$ are only used when 
it is impossible to achieve the fluent $f$ in a strongly or weakly compliant manner, respectively.

The combination of compliance optimization with the first three behavior modes allows for more possible configurations.
For example, adding optimization to the \textit{subordinate} option makes a \textit{cautious} mode.
In this mode, the agent will try to mimic the behavior of the \textit{paranoid} mode (all strongly compliant actions), 
but ultimately it will reduce to \textit{subordinate} (all weakly compliant actions) in the worst case.
Likewise, adding optimization to the \textit{utilitarian} mode adds two options: \textit{best effort} and 
\textit{subordinate when possible}.
\textit{Best effort} prefers strong compliance over weak compliance and weak compliance over non-compliance, 
but ultimately permits non-compliance when no better alternatives exist.
\textit{Subordinate when possible} prefers weak compliance over non-compliance but does not optimize from weak 
compliance to strong compliance.

\smallskip
A new feature of this approach to optimization is the ability to optimize within the weakly and non-compliant categories.
Consider two weakly compliant activities, 1 and 2, where activity 1 has more weakly compliant actions than activity 2.
Since weakly compliant actions do require a concurrent $ignore\_not\_permitted(a)$ action, 
activity 1 will have more $ignore\_not\_permitted(a)$ actions than activity 2.
Hence, activity 2 will be preferred to activity 1, even though they both fall in the weakly compliant category.
Gelfond and Lobo \cite{gl08} do not consider such a feature.


\medskip
\noindent
{\bf Obligation Policies and Agent Behavior. }
So far we discussed authorization policies induced by an \aopl policy.
To address obligation policies, we add policy fluents $obl\_compliant(do\_action)$ and 
$obl\_compliant(refrain\_from\_action)$ with policy actions $ignore\_obl(a)$ and $ignore\_obl(neg(a))$, 
as seen in Table \ref{table:policy_fluents_and_actions}.
(For configurability, we consider obligation policies to {\em do} actions and to {\em refrain} from actions separately).
We extend the definition of $policy\_compliant(f)$ to require both $obl\_compliant(do\_action)$ and 
$obl\_compliant(refrain\_from\_action)$ to be true.
Like with authorization compliance, if $obl(a)$ is true but action $a$ does not occur, 
then $obl\_compliant(do\_action)$ becomes false and remains false by inertia.
Likewise for $obl\_compliant(refrain\_from\_action)$.
If $ignore\_obl(a)$ or $ignore\_obl(neg(a))$ are performed, then these effects on the $obl\_compliant$ 
fluents are temporarily waived.

\begin{table}[!hbt]
    \centering
		 \begin{tabular}{ l | l | l | l }
        \hline\hline
        & {\bf Honor $obl(\neg a)$} & {\bf Prefer honoring $obl(\neg a)$} & {\bf Ignore $obl(\neg a)$} \\
        \hline
        {\bf Honor $obl(a)$} & Subordinate & Permit commissions & (Not reasonable) \\
        \hline
        {\bf Prefer honoring $obl(a)$}  & Permit omissions  & Best effort & (Not reasonable) \\
        \hline
        {\bf Ignore $obl(a)$} & (Not reasonable) & (Not reasonable) & Utilitarian \\
        \hline\hline
    \end{tabular}
    \caption{\apia Obligation Policy Modes}
    \label{table:apia_obligation_modes}
\end{table}

There are five different configurations (or behavior modes) 
an agent in the \apia architecture can have regarding its obligation policy 
(see Table \ref{table:apia_obligation_modes}).
When in \textit{subordinate} mode, the agent will never use either $ignore\_obl(a)$ and $ignore\_obl(neg(a))$ actions.
Hence, all activities achieving $policy\_compliant(f)$ will be compliant with both aspects of its obligation policy.
When in \textit{best effort} mode, the agent prefers using other actions over these policy actions.
Hence, activities will be compliant if possible but may include non-compliant elements when no other goal-achieving activities exist.
The \textit{permit omissions} and \textit{permit commissions} options are variations of these modes.
Mode \textit{permit omissions} is like \textit{best effort} with regards to $obl(a)$ policy statements, but like \textit{subordinate} with regards to $obl(\neg a)$ policy statements.
Likewise, \textit{permit commissions } is like \textit{subordinate} with regards to $obl(a)$ policy statements but like \textit{best effort} regarding $obl(\neg a)$ statements.
\textit{Utilitarian} mode, like with authorization policies, reduces the behavior of an \apia agent with respect to its obligation policy to that of an \aia agent.

\medskip
\noindent
{\bf Behavior Mode Configurations. }
An agent's combined authorization and obligation policy configuration can be represented by a 2-tuple $(A, O)$, 
where $A$ is the authorization mode and $O$ is the obligation mode.
When an \apia agent is running in mode $(utilitarian, utilitarian)$, its behavior 
reduces to that of an \aia agent (i.e., policy actions are not used in this mode).
This is due to an optimization we provide internally.

For each \apia configuration, we adjust the definition of 
$policy\_compliant(f)$ such that excess policy actions are not required.
For instance, in the case of an agent with a subordinate authorization mode, 
we adjust $policy\_compliant(f)$ such that $ignore\_not\_permitted(a)$ is 
never needed since such an agent always disregards strong compliance.


\section{Examples}

To demonstrate the operations of an agent in the \apia architecture, we will introduce a series of examples
that illustrate prototypical cases.
For conciseness, we will focus on three \apia configurations: {\em (paranoid, subordinate)}, 
{\em (best effort, best effort)}, and {\em (utilitarian, utilitarian)},
and we limit ourselves to examples about authorization policies.

\subsection{Example A: Fortunate case}

To begin with a simple case, suppose that two people are in an office space that has four rooms with doors in between them.
Room 1 is connected by door $d_{12}$ to Room 2.
Room 2 is connected by door $d_{23}$ to Room 3 and so on.
Door $d_{34}$ has a lock and is currently in the unlocked position. 
Suppose our agent, Alice, wants to greet another agent, Bob.
%
This scenario is represented by a dynamic domain description
that considers:
\begin{itemize}
\item {\bf Fluents:} $door\_locked(D)$ for each door $D$, $in\_room(P, R)$, $greeted\_by(P, A)$ where person $P$ is greeted by person $A$; and
\item {\bf Actions:} $move\_through(A, D)$, $lock\_door(A, D)$, $unlock\_door(A, D)$, $greet(A, P)$, 
where $A$ is the person doing the action, $D$ is a door, and $P$ a person (the direct object of the action).
\end{itemize}

Assume that agent Alice is given a policy specifying that all actions are permitted
along with the following pre-computed activity that is stored in her knowledge base as the set of facts:
$$
\begin{array}{l}
		\{activity(1), \ \ \ \ length(1, 4), \ \ \ \ goal(1, policy\_compliant(greeted\_by(alice, bob))),\\
    \ \ component(1, 1, move\_through(alice, d_{12})), \ \  component(1, 2, move\_through(alice, d_{23})), \\ 
		\ \ component(1, 3, move\_through(alice, d_{34})), \ \ component(1, 4, greet(alice, bob))\}\\
\end{array}
$$

Before the control loop begins, Alice observes that she is in Room 1, Bob is in Room 4, the door $d_{34}$ is unlocked, and that she has not yet greeted Bob.

At timestep 0, the first iteration of the control loop begins.
In this first step, Alice analyzes her observations and interprets unexpected observations by assuming undetected exogenous actions occurred.
None of her observations are unexpected, so no exogenous actions are assumed to occur.
Alice then intends to wait at timestep 0.
Alice attempts wait.
Alice observes that her wait action was successful and that, in the meantime, the exogenous action $$select(policy\_compliant(greeted\_by(alice, bob)))$$ happened.
The time step is incremented and Alice does not observe any fluents.

The second iteration of the control loop begins.
Alice analyzes her observation of \\ $select(policy\_compliant(greeted\_by(alice, bob)))$ and determines that \\ $active\_goal(policy\_compliant(greeted\_by(alice, bob)))$ is true.
Alice then starts planning to achieve $policy\_compliant(greeted\_by(alice, bob))$ and determines that she intends to start activity 1.
Since each action in activity 1 is strongly compliant, no policy actions are needed.

The rest of the example is very straight forward and is almost identical to scenarios discussed by \cite{bj13,bgb14} in the \aia architecture.

\subsection{Example B: Strong compliance degrades to weak compliance}
Let us consider a less fortunate example, in which a strongly compliant activity becomes weakly compliant 
due to an unexpected environmental observation.
In the same scenario, suppose we modify Alice's policy from Example A such that regarding $greet(A, P)$ we have:
\begin{equation}
\label{eq:alice_policyB}
\begin{array}{lll}
    permitted(greet(A, P)) & \textbf{ if } & \neg busy\_working(P) \\
    permitted(greet(A, P)) & \textbf{ if } & busy\_working(P), \ \ in\_room(P, R), \\
            & & door\_connects(D, R), \ \ knocked\_on\_door(D)
\end{array}
\end{equation}
We also have new fluents $busy\_working(P)$ and $knocked\_on\_door(D)$, and actions 
$begin\_working(A)$ and $knock\_on\_door(A, D)$.
Let Alice's knowledge base contain two additional activities, 2 and 3, with the same goal as 1
and defined by the sets of facts:
$$
\begin{array}{l}
		\{activity(2), \ \ \ \ length(2, 5), \ \ \ \ goal(2, policy\_compliant(greeted\_by(alice, bob))),\\
    \ \ \ \ component(2, 1, move\_through(alice, d_{12})), \ \  component(2, 2, move\_through(alice, d_{23})), \\ 
		\ \ \ \ component(2, 3, knock\_on\_door(alice, d_{34})), \ \ component(2, 4, move\_through(alice, d_{34})), \\ 
		\ \ \ \ component(2, 5, greet(alice, bob)),\\
		\ \ activity(3), \ \ \ \ length(3, 2), \ \ \ \ goal(2, policy\_compliant(greeted\_by(alice, bob))),\\
    \ \ \ \ component(3, 1, knock\_on\_door(alice, d_{34})), \ \ component(3, 2, move\_through(alice, d_{34}))\}
\end{array}
$$

Alice observes that Bob is not busy working, in addition to the initial observations of Example A.
At timestep 0, the first iteration of the control loop begins.
During the second iteration of the control loop (at timestep 1), Alice plans to achieve the $policy\_compliant(greeted\_by(alice, bob))$ goal.
Since she believes Bob is not busy working, activity 1 is still strongly compliant and so is activity 2.
Alice chooses activity 1 over activity 2 because it requires a shorter sequence of actions.
She then executes activity 1 like in Example A until she enters Room 3, at which points she observes that Bob is busy working.
During the next iteration (at timestep 5), the agent interprets this observation by inferring that $begin\_working(bob)$ happened at the previous timestep (4).

As a result, activity 1 becomes weakly compliant.
Since Bob is busy working but Alice has not knocked on the door, no policy statement describes whether our next action, 
$greet(alice, bob)$, is compliant or not.
If Alice is operating in {\em (utilitarian, utilitarian)} mode, she continues the execution of activity~1 and greets Bob anyway.
(This happens without the use of policy actions due to our internal optimizations).
Otherwise, Alice will stop the activity and then either refuse to plan another weakly compliant activity or use a concurrent policy action to dismiss this event.

If our agent is running in {\em (paranoid, subordinate)}, Alice will refuse to execute a weakly compliant activity.
Through planning, Alice will discover that a new activity that includes knocking at the door is strongly compliant (e.g.~activity 3) and begin its execution.
If our agent is running in {\em (best effort, best effort)}, she will behave likewise because activity 3 is strongly compliant.
The difference is that, if there did not exist a strongly compliant activity, she would plan a new activity that involved a policy action and greeted Bob anyway.
Alice knocks on door $d_{34}$ at timestep 7, greets Bob at timestep 8, and stops activity 3 at timestep 9.

\subsection{Example C: Compliance degrades to non-compliant}

Suppose we take policy rule (\ref{eq:alice_policyB}), make it defeasible, and add this \aopl rule :
$$
\begin{array}{lll}
    \neg permitted(greet(A, P)) & \textbf{ if } & busy\_working(P), \ \ supervisor\_to(P, A)
\end{array}
$$

Now, let us imagine that Bob is Alice's supervisor.
Similar to Example B, our agent executes activity 1 until the observation that Bob is busy working.
This time, we have a strict authorization statement forbidding greeting Bob since he is Alice's supervisor.
Under the {\em (utilitarian, utilitarian)} option, we proceed on with activity 1 anyway.
With the {\em (paranoid, subordinate)} option, our agent stops activity 1 but 
cannot construct a new activity that achieves the goal subject to its policy.
Hence, the goal is futile and the agent waits until its environment changes such that a strongly 
compliant activity exists.
Under the {\em (best effort, best effort)} option however, our agent constructs a new activity that contains greets Bob anyway.
The activity contains: $greet(alice, bob)$, $ignore\_not\_permitted(greet(alice, bob))$, and $ignore\_neg\_permitted(greet(alice, bob))$.

\subsection{Example D: Hierarchy of contradictory defeasible statements}

Further extending Example C, suppose we turn all policy statements from Example A into defeasible ones 
(i.e., all actions are {\em normally} permitted) and add another policy statement:
$$
\begin{array}{lllll}
    m1(A, D): & \textbf{ normally } & permitted(move\_through(A, D)) & & \\
    m2(A, D): & \textbf{ normally } & \neg permitted(move\_through(A, D)) & \textbf{ if } & in\_room(A, R_1), \\
        & & & & door\_connects(D, R_2),\\ 
				& & & & private\_office(R_2),\\ 
				& & & & R_1 != R_2
\end{array}
$$
and the static $private\_office(R)$ with $private\_office(r_4)$ as a fact.
Since we have two contradictory defeasible statements, we need to add a preference between the two (without a preference our agent can non-deterministically choose between which of the two rules to apply).
If we add:
$$prefer(m2(A, D), m1(A, D))$$
then, when Alice observes that Bob is not busy working at the beginning of the scenario, an agent running in
 {\em (paranoid, subordinate)} mode will immediately consider the goal to be futile.
Unlike in Example C, our agent knows this immediately because $private\_office$ is a static, not an unexpected observation.
If our agent is running in {\em (best effort, best effort)} mode, it creates an activity like activity 1, 
except that it contains $ignore\_not\_permitted(greet(alice, bob))$ and $ignore\_neg\_permitted(greet(alice, bob))$.
Our utilitarian agent, like always, completely ignores our policy and executes activity 1.


\section{Implementation}

In this section, we discuss two important implementation aspects:
the refactoring of the \aia implementation including its Theory of Intentions and control loop,
and the implementation of the \apia control loop using \textsc{clingo}'s Python API.

\subsection{\texorpdfstring{\aia}{AIA} Theory of Intentions and Control Loop}

Since \apia takes \aia as a basis, we first update Blount {\em et al.}'s \cite{bgb14} \aia implementation 
such that it requires a state-of-the-art solver: \textsc{clingo} (version 5.4.1).\footnote{
    Our updated \aia implementation is available at \url{https://gitlab.com/0x326/miami-university-cse-700-aia.git} and is released under the MIT open-source license.
} 
For this purpose, we re-implement the \aia logic program in ASP using only the description of the architecture presented by 
Blount {\em et al.} \cite{bj13,bgb14}. 
During this process, we make minor modifications to \aia as a whole.
First, we refactor the arrangement of ASP rules into multiple files according to their purpose in the \aia architecture (e.g.~whether they are part of the Theory of Intentions, \aia's rules for computing models of history, or the \aia intended action rules).
Second, we refactor the names of mental fluents in the Theory of Intentions so that their names are more descriptive and self-documenting.
Thirdly, we extensively add inline comments to each ASP rule with reference quotations and page numbers from 
Blount {\em et al.}'s work.
Lastly, we make minor corrections to ASP rules to match the translation of particular scenarios (i.e., histories)
with the mathematical definitions proposed by Blount {\em et al.}.

In addition to upgrading the \aia logic program, we also refactor the implementation of the \aia control loop.
In his dissertation, Blount \cite{bj13} introduced the \aia Agent Manager.
This is an interactive Java program that allows an end-user to assign values to agent observations in a graphical interface for each control loop iteration.
Since this requires manual input, it does not easily lend itself to automation and reproducibility of execution, which are required for performance benchmarking.
Furthermore, the \aia Agent Manager is structured around interacting with an underlying solver using subprocesses and process pipes.
While the \aia Agent Manager could conceivably invoke \textsc{clingo} as a subprocess, \textsc{clingo} 5 provides a unique opportunity for more advanced integrations using its Python API.

Because of these two points, we replace the \aia Agent Manager with a new implementation of the \aia control loop written in Python 3.9.0.
This new implementation uses a command-line interface and allows for reproducible execution through ASP input files.
Since this control loop is also the basis for our \apia implementation, we will discuss it more in the next subsection.

\subsection{Python Component}
\label{subsec:apia_control_loop}

We provide an implementation of the \aia control loop for the \apia architecture.\footnote{
    Our \apia implementation is available at \url{https://gitlab.com/0x326/miami-university-cse-700-apia.git} and is released under the MIT open-source license.
}
The \apia control loop is implemented using Python 3.9.0 and \textsc{clingo} 5.4.1 using \textsc{clingo}'s Python API.
We provide two modes: an automatic mode and a manual mode.
The automatic mode is intended to be used for normal execution while the manual mode is intended to aid in debugging unexpected output in answer sets.
The automatic mode uses a command-line interface to specify the ASP files of the input domain, the observations of the agent, and the \apia policy compliance mode the agent should use.
The control loop then provides human-readable output as to what happens at each control loop step 
(see Figure \ref{fig:apia} in the appendix).


In the case of unexpected output, the manual mode allows one to examine the answer set at each step of the control loop.
It also provides scripts to highlight differences between answer sets of different timesteps in a visual manner and to step through the control loop like one would do in a traditional debugger.
Additionally, manual mode addresses certain violations of \aia and \aopl underlying assumptions.
For example, it generates an invalid predicate when there exists an action that is neither a physical, mental, or policy action.
Likewise when an action is neither an agent action nor an exogenous action.
In addition, it generates an invalid predicate when an \aopl policy statement describes an object that is not declared as an action.
These rules have been very useful in debugging the implementation of the \apia architecture and they will aid future end-users who encode and execute scenarios using this architecture.
Since these rules are intended during debugging, they are not executed during the automatic mode.


\section{Conclusions and Future Work}
\label{sec:conclusions}

In this paper, we created an architecture for a policy-aware intentional agent by
bridging together previous work on intentional agents \cite{bj13,bgb14}
and reasoning algorithms for authorization and obligation policies \cite{gl08}.
A main difficulty was adapting the work on policy compliance so that
it would be relevant for an agent deciding on which course of actions to take. 
While Gelfond and Lobo's work could determine whether a trajectory (i.e., sequence of actions)
was strongly compliant, weakly compliant, or non-compliant, we introduced a wider range of
agent behavior modes, which additionally explore the interactions between authorization and obligation policies.

This work can be further expanded by refining the decision making process in the planning phase of \apia
by introducing a relative ranking system between activities that would achieve the same goal,
based on the number of actions that are strongly, weakly, or non-compliant. 
Moreover, it would be interesting to allow the agent's controller to switch behavior modes while the agent
is active, in the middle of executing an activity.


\bibliographystyle{eptcs}
\bibliography{apia}

\newpage
\appendix
\section{\texorpdfstring{\apia}{APIA} Output}

\begin{figure}[!hbt]
    \centering
    \includegraphics[width=0.95\textwidth]{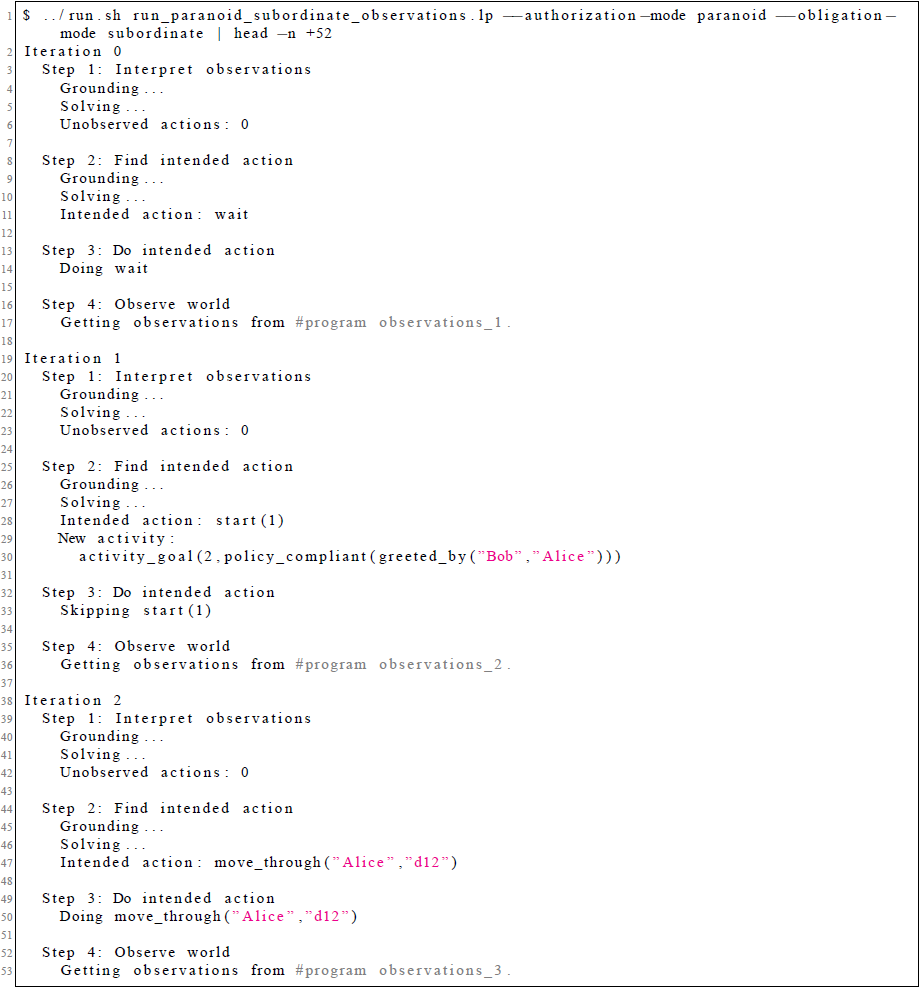}
    \caption{Automatic execution of Example A using configuration $(paranoid, subordinate)$}
    \label{fig:apia}
\end{figure}

\label{lastpage}

\end{document}